\def\ber{\begin{eqnarray}}
\def\eer{\end{eqnarray}}
\def\beq{\begin{equation}}
\def\eeq{\end{equation}}

\documentclass[amsmath,amssymb,nofootinbib]{revtex4}
\usepackage[dvips]{graphicx}

\begin{document}

\title{ACCELERATING COSMOLOGY IN RASTALL'S THEORY}

\author{Monica Capone}
\email{monica.capone@unito.it}
\affiliation{Dipartimento di Matematica,  Universit\`a di Torino, Via Carlo Alberto 10, 10125 - Torino, Italy\\
 INFN, Sezione di Torino, Via Pietro Giuria 1, 10125 - Torino, Italy}

\author{Vincenzo Fabrizio Cardone}
\email{winnyenodrac@gmail.com}
\affiliation{Dipartimento di Fisica Generale ``Amedeo Avogadro'', Universit\`a di Torino, 
Via Pietro Giuria 1, 10125 - Torino, Italy\\ INFN, Sezione di Torino, Via Pietro Giuria 1, 10125 - Torino, Italy}

\author{Matteo Luca Ruggiero}
\email{matteo.ruggiero@polito.it}
 \affiliation{UTIU, Universit\`a Telematica Internazionale Uninettuno, Corso Vittorio Emanuele II 39, 00186 - Roma, Italy \\ Dipartimento di Fisica, Politecnico di Torino, Corso Duca degli Abruzzi 23, 10129 - Torino, Italy \\ INFN, Sezione di Torino, Via Pietro Giuria 1, 10125 - Torino, Italy}

\date{\today}

\begin{abstract}
In an attempt to look for a viable mechanism leading to a present-day accelerated 
expansion, we investigate the possibility that the observed cosmic speed up may be 
recovered in the framework of the Rastall's theory, relying on the non-conservativity 
of the stress-energy tensor, i.e. $T^{\mu}_{\nu ; \mu} \neq 0$. We derive the 
modified Friedmann equations and show that they correspond to Cardassian-like equations.
We also show  that, under suitable assumptions on the equation
of state of the matter term sourcing the gravitational field, it is indeed possible to 
get an accelerated expansion, in agreement with  the Hubble diagram of 
both Type Ia Supernovae (SNeIa) and Gamma Ray Bursts (GRBs). Unfortunately, to achieve such
a result one has to postulate a matter density parameter much larger than the typical 
$\Omega_M \simeq 0.3$ value inferred from cluster gas mass fraction data.
\end{abstract}
\maketitle


\section{Introduction}\label{sec:intro}

The observed cosmic speed up \cite{Riess98,Perlmutter99,Bennet03} questions the validity of General Relativity (GR)
on large scales. In fact, if on one hand the model of gravitational interaction as described by Einstein's theory
is in agreement with  many observational tests on relatively small scales, as Solar System and binary pulsars
observations show \cite{Will06}, it is well known that in order to make GR agree with the observed  acceleration of the Universe the existence of \textit{dark energy}, a cosmic fluid having exotic properties, has been postulated. Actually, many candidates for explaining the nature
of dark energy have been proposed (see e.g. \cite{peebles03}, \cite{dark} and references therein), some of them relying on the modification of
the geometrical structure of the theory, some others  on the introduction of physically (up to day) unknown fluids into the equations governing the behaviour of our universe.  Moreover, it is interesting to point out that the problem of explaining the acceleration of the Universe has been addressed also in the framework of GR (see \cite{kolb} and references therein).

In this context, we want to consider here a generalization of  Einstein's  theory, the so-called Rastall's model \cite{rastall}, based on the requirement 
that  the stress-energy tensor for the matter/energy content is not conserved, i.e. 
$T^{\mu}_{\nu ; \mu} \neq 0$.  Rastall's model has been initially motivated by the need for a theory able to 
allow a non-conservativity of the source stress-energy tensor without 
violating the Bianchi identities. As such, the original theory was based on
purely phenomenological motivations and directly started with the field equations
without any attempt to derive them from a variational principle (even if there have been subsequent attempts to deduce Rastall's field equations from a variational principle,  but none of them have succeeded \cite{smalley93,lind82}).

As for the confrontation with the data, it is interesting to point out that Rastall's field equations, in vacuum, are equivalent to GR ones: as a consequence, all classical tests of GR are correctly reproduced. On the other hand, it could be useful to test the cosmological predictions of the theory, by considering the solutions within the cosmological fluid. Our work is motivated by the fact that Rastall's theory was introduced more than 30 years ago, so it is interesting to test it against the recent cosmological data. 
In particular, we focus on the possibility of describing the accelerated expansion of the Universe in Rastall's framework, by investigating the conditions that the parameters of the theory have  to fulfill in order to reproduce the data. Furthemore,  prompted by a recent paper \cite{Raw}, we check whether the Cardassian model \cite{Card,fay06} can be derived by Rastall's tensorial equations, because, despite the fact that this model passed almost all observational tests, it is purely phenomenological.

The plan of the paper is as follows: we will firstly give an introduction 
to Rastall's model in Sect. \ref{sec:rastall}, while the corresponding cosmological scenario and the analogy with 
the Cardassians expansion model is worked out in Sect. \ref{sec:cardass}. In order to check the possible viability
of the Rastall's proposal, we test the model with respect to the SNeIa and GRBs Hubble diagram, as detailed 
in Sect. \ref{sec:data}. Conclusions are finally 
presented in Sect. \ref{sec:conc}. 

\section{Rastall's Model} \label{sec:rastall}

In 1972 P. Rastall \cite{rastall} explored a model in which the stress-energy tensor of the source 
of the gravitational field, $T_{\mu \nu}$, was not conserved, i.e. the condition $T^{\mu}_{\nu ; \mu} \neq 0$
is imposed a priori. 

Indeed, Einstein equations\footnote{Throughout the paper, spacetime is assumed to have the signature
$(+, -, -, -)$, and Greek indices run from 0 to 3.} read

\beq
G_{\mu \nu}\doteq R_{\mu \nu}- \frac 1 2 g_{\mu \nu}R=\kappa_{GR} T_{\mu \nu}\ ,\ \label{eq:Eeq}
\eeq
where the Ricci tensor is obtained from a metric connection, so that $R_{\mu\nu}=R_{\mu\nu}(g)$ and  the scalar
curvature $R$ has to be intended as $R\equiv R(g)
=g^{\alpha\beta}R_{\alpha \beta}(g)$; furthermore we have set $\kappa_{GR} = \frac{8 \pi G}{c^{4}}$.

These equations naturally imply the stress-energy tensor conservation as a consequence of the contracted Bianchi identities,
\beq
G^{\mu}_{\nu ; \mu}=0\ .\ \label{eq:Bianchi}
\eeq 
It is therefore worth wondering whether it is possible to fulfill the requirement $T^{\mu}_{\nu ; \mu} \neq 0$
without violating Eqs.(\ref{eq:Bianchi}). A possible way out could be introducing further geometrical 
terms on the right hand side of Einstein equations, even if one should ask whether this makes sense. Actually,
if we insist in  deriving these relations from a metric variational approach, the sudden answer would 
be of course negative: in this case the stress-energy tensor would be surely conserved by construction, 
so no way to escape the conditions $T^{\mu}_{\nu ; \mu} = 0$. 

Another remark against the non-conservativity focuses on the equivalence principle: as a 
matter of fact, the conservation of the stress-energy tensor is tested with high accuracy 
in the realm of Special Relativity (SR). Then, one jumps to the realm of GR just invoking the principle of minimal coupling. However, one has to go easy with such an approach, 
as this principle could be misleading \cite{Traut}. To give an example, when passing from GR
to SR, we completely miss the information provided by terms explicitly depending on the curvature 
tensor, $R^{\alpha}_{\mu \beta \nu}$, as it becomes identically zero when the spacetime becomes flat. This 
means that the two sets of equations

\beq
\nabla_{\alpha}j^{\alpha} = 0
\eeq
and

\beq
\nabla_{\sigma}j^{\sigma}+R^{\alpha}_{\mu \beta \nu} \nabla_{\alpha}j^{\mu} \nabla^{\beta}j^{\nu} = 0 \ ,
\eeq
give exactly the same equations, i.e. $\partial_{\alpha}j^{\alpha}=0$, in SR. So, the straightforward application of the equivalence principle in writing conservation laws should be carefully considered.

The question is now how to pick up a proper geometrical term such that the Bianchi 
identities are still valid, but nevertheless the conservation of the stress-energy
tensor of the gravity source is violated. To resume, we ask for a four-vector, say $a_{\nu}$,
such that (i)  $T^{\mu}_{\nu ; \mu} = a_{\nu}$; (ii) $a_{\nu} \neq 0$ on curved spacetime, but
$a_{\mu} = 0$ on flat spacetime in order not to conflict with the validity of SR. Both these 
properties hold for the Rastall's proposal, that is 

\beq
T^{\mu}_{\nu ; \mu} = \lambda R_{,\nu}\ ,\ \label{eq:Rast req}
\eeq
$\lambda$ being a suitable non-null dimensional constant.

Because of the assumption (\ref{eq:Rast req}), the field equations are obviously modified and now read

\beq
R_{\mu \nu}- \frac{1}{2} \left ( 1 - 2 \kappa_{r} \lambda \right ) R g_{\mu \nu} 
= \kappa_{r} T_{\mu \nu} \ , 
\label{eq:rast1}
\eeq 
where $\kappa_{r}$ is a dimensional constant to be determined in order to give the right Poisson 
equation in the static weak-field limit. It is manifest that in vacuum, where $T_{\mu\nu}=0$, Rastall's field equations (\ref{eq:rast1}) are equivalent to GR ones. 

As a matter of fact, the same set of equations can be obtained as the result of guesswork, that 
is  assuming the left hand side of the sought after equations to be a symmetric tensor only
consisting of terms that are linear in the second derivative and/or quadratic in the first 
derivatives of the metric \cite{Wein}. Moreover, the time-time component of such equations  
must give the Poisson equations back for a stationary weak-field. Accordingly, the only requirement we drop with respect to the derivation of 
the Einstein equations is the one concerning the conservation of $T_{\mu \nu}$. Hence, starting 
from $G_{\mu \nu} = C_{1} R_{\mu \nu} + C_{2} R g_{\mu \nu}$, with $C_{1}$ and $C_{2}$ appropriate 
constants, we end up with Eqs.(\ref{eq:rast1}) again, provided that we set:

\begin{eqnarray}
C_{2} & = & \frac {C_{1} \left( C_{1} - 2 \right )}{2 \left ( 3 - 2 C_{1} \right ) } \ ,  \\
\kappa_{r} & = & \frac {8 \pi G}{C_{1} c^{4}} \equiv \frac{\kappa_{GR}}{C_{1}} \ ,  \label{eq:kappar} \\ 
\kappa_{r} \lambda & = & \frac{C_{1} \left( 1 - C_{1} \right)}
{2 \left ( 3 - 2 C_{1} \right )} \ , \label{eq:kappar lambda}
\end{eqnarray}
where we have chosen to rewrite all the other constants in terms of the $C_{1}$. Note, in particular, that 
the coupling constant between matter and geometry, $\kappa_{r}$, is not the same as in GR, unless $C_{1} 
= 1$, that is $\lambda = 0$ (i.e., we consistently go back to GR).

Taking the trace of Eqs.(\ref{eq:rast1}) gives us the structural or master equation \cite{Ferr}\,:
\beq
\left ( 4 \kappa_{r} \lambda - 1 \right ) R = \kappa_{r} T \ .  
\label{eq:mast rast}
\eeq
For a traceless stress-energy tensor, $T = 0$ (as for the electromagnetic tensor) and two possibilities
arise. The first is that $R = 0$ so that we get no differences with standard GR. On the other hand, one could also
solve Eq.(\ref{eq:mast rast}) setting $\kappa_r \lambda = 1/4$, whatever the value of $R$ is. However, 
inserting this condition in Eq.(\ref{eq:kappar lambda}), we get a complex value for $C_1$ which is 
clearly meaningless. Therefore, we hereafter assume that $\kappa_r \lambda \neq 1/4$.

Another fundamental question concerns geodesic motions. As it is well known, the equations $T^{\mu}_{\nu ; \mu}
= 0$ are nothing but the equations of motion of the fluid we are dealing with. The problem is then, 
what sort of curves are described in a curved spacetime by a fluid whose stress-energy tensor is not 
conserved. Following the calculations made by Rastall, we find that in his model geodesics are those 
curves characterized by the fact that the scalar curvature $R$ is constant along them. Moreover, it is still 
possible to speak of conservation of energy for an ideal fluid \cite{rastall}, but again provided that 
$R$ is constant along the time-like four-velocity vector of the fluid, $u_{\mu}$. The question 
remains whether particles creation takes place in the regions where this condition does not hold.

It is worth mentioning that the Rastall's equations (\ref{eq:rast1}) can be recast into the same form 
as the usual Einstein ones. Indeed, one can immediately write

\beq
G_{\mu \nu}=\kappa_{r} S_{\mu \nu}\ ,\ \label{eq:rast S}
\eeq
where 

\beq
S_{\mu \nu} = T_{\mu \nu} - \frac {\kappa_{r} \lambda}{4\kappa_{r} \lambda - 1} g_{\mu \nu} T \ . \
\label{eq:newten}
\eeq
By construction, this new stress-energy tensor is conserved, $S^{\mu}_{\nu ; \mu} = 0$. On introducing $S_{\mu \nu}$, we can recover all the known solutions of Einstein GR by simply 
taking care of the difference between $S_{\mu \nu}$ and $T_{\mu \nu}$. Furthermore, if we assume 
$T_{\mu \nu} = (\rho + p) u_{\mu} u_{\nu} - p g_{\mu \nu}$, i.e. the source is a perfect fluid with
energy density $\rho$ and pressure $p$, we can explicitely work out an expression for $S_{\mu \nu}$. 
This turns out to be still a perfect fluid, provided we redefine its energy density and pressure as

\begin{eqnarray}
\rho_S & = & \frac{(3 \kappa_r \lambda - 1) \rho + 3 \kappa_r \lambda p}
{4 \kappa_r \lambda - 1} \ , \\
p_S & = & \frac{\kappa_r \lambda \rho + (\kappa_r \lambda - 1) p}
{4 \kappa_r \lambda - 1} \ .
\end{eqnarray}
In order to obtain the value of the coupling constant $\kappa_r$, we remember that
the time-time component of the modified equations should recover the Poisson equation
in the static weak-field limit. One thus gets:

\beq
\frac {\kappa_{r}}{4 \kappa_{r} \lambda - 1 } 
\left ( 3 \kappa_{r} \lambda - \frac{1}{2} \right ) = \kappa_{GR} \ ,
\label{eq: kappagrrast}
\eeq
whence it is immediate to derive exactly the same coupling as in Einstein gravity only when 
$\lambda = 0$, that is when the conservation of $T_{\mu \nu}$ is granted.

\section{A Cardassian analog and the cosmic speed up}\label{sec:cardass}

It has been recently claimed \cite{Raw} that a Cardassian-like \cite{Card,fay06} modification of the 
Friedman equation in the form

\beq
H^{2} = \frac{8 \pi G}{3c^{2}} \left [ 
\rho + B(t)(\rho - 3p)^{n} \right ] \ ,
\label{eq:Taha}
\eeq
can be obtained from Rastall-like equations, where $B(t)$ is a function of the cosmic time $t$. We would now like to show that, although it is 
indeed possible to recast the Rastall's theory equations in such a way that a Cardassian-like model
is recovered, the parameter $B$ in (\ref{eq:Taha}) must be a constant. 

To this aim, we derive the cosmological equations for the Rastall's theory. We first remember that,
when the isotropic and homogenous Robertson-Walker (RW) metric is adopted, in GR one gets the 
usual Friedmann equations 

\begin{eqnarray}
H^2 & = & \frac{\kappa_{GR} c^2}{3} \rho \ , \label{eq:FRW1} \\
\dot{H} + H^2 & = & - \frac{\kappa_{GR} c^2}{6} \left ( \rho + 3 p \right ) \ , \label{eq:FRW2}
\end{eqnarray}
where $H = \dot{a}/a$ is the Hubble parameter, $a$ the scale factor and a dot denotes the derivative with respect to 
cosmic time $t$. To get the corresponding equations for the Rastall's theory, one has to insert
the RW metric into Eqs.(\ref{eq:rast1}) and consider the only independent equations that can be obtained, that is
\begin{eqnarray}
3 \dot{a}^{2} - 6 \kappa_{r} \lambda \left ( \dot{a}^{2} + a \ddot{a} \right ) 
& = & \kappa_{r} a^{2} \rho \ , \label{eq:titi} \\
\dot{a}^{2} + 2 a \ddot{a} + 6 \kappa_{r} \lambda \left ( \dot{a}^{2} + a \ddot{a} \right )
& = & - \kappa_{r} a^{2} p \ , \label{eq:esseesse}
\end{eqnarray}
respectively. The master equation thus becomes 

\beq
6 \left ( \dot{a}^{2} + a \ddot{a} \right ) = - \kappa_{r} a^{2} \left ( \rho - 3 p \right) \ , \label{eq:masterfrw}
\eeq
so that multiplying Eq.(\ref{eq:esseesse}) by $-3$ and then adding to Eq.(\ref{eq:titi}) we finally get the first 
modified Friedmann's Equation\,:

\beq
\frac{\ddot{a}}{a} = \frac{\kappa_{r}}{6} 
\frac{\rho + 3 p - 6 \kappa_r \lambda (\rho + p)}{4 \kappa_r \lambda - 1}\ ,
\label{eq:FR R1}
\eeq
which makes it possible to directly infer the sign of the acceleration. To obtain the second 
modified Friedmann's equation, it is easier to proceed in a slightly different way. Let us first 
take the Rastall's equations in the form

\beq
R_{\mu \nu} = \kappa_{r} \left ( T_{\mu \nu} - \frac 1 2 
\frac{2 \kappa_{r} \lambda - 1}{4 \kappa_{r} \lambda - 1} T g_{\mu \nu} \right ) \ .
\eeq
By inserting the RW metric and adding up Eq.(\ref{eq:titi}) with three times Eq.(\ref{eq:esseesse}),
we eventually obtain

\beq
H^2 = \frac{\kappa_{r}}{6} \left [ (\rho + 3p) + 
\frac{2 \kappa_{r} \lambda - 1}{4\kappa_{r} \lambda -1}(\rho -3p) \right ] \ . 
\label{eq:FR R2}
\eeq
It is then only a matter of algebra to rearrange Eqs.(\ref{eq:FR R1}) and (\ref{eq:FR R2}) to
write them as

\beq
H^2 = \frac{\kappa_{r}}{3} \left [ \rho - \frac{\kappa_{r} \lambda}{4 \kappa_{r} \lambda - 1}
(\rho -3 p) \right ] \ , 
\label{eq: rf1}
\eeq

\beq
\dot{H} = \frac{\kappa_{r}}{2 ( 4\kappa_{r} \lambda - 1)}
\left [ \rho + p - 4\kappa_{r} \lambda (\rho - p) \right ] \ ,
\label{eq: rf2}
\eeq
which reduce to the standard Friedmann equations (\ref{eq:FRW1}) and (\ref{eq:FRW2}) when the 
parameter $\lambda$ is switched off. Note also that Eq.(\ref{eq: rf1}) has indeed the same
expression as the Cardassian-like Eq.(\ref{eq:Taha}) provided we set $n = 1$ and accordingly
redefine the parameter $B$. However, it is straightforward to show that $B$ must be a constant. Indeed,
from the condition 

\beq
S^{\mu}_{\nu ; \mu}= 0 \ , 
\label{eq:conserv}
\eeq
it is immediate to demonstrate that $\kappa_{r} \lambda$ must be a constant by simply inserting the 
master equation (\ref{eq:masterfrw}) into Eq.(\ref{eq:conserv}) and using the Rastall's requirement $T^{\mu}_{\nu ; \mu} =
\lambda R_{,\nu}$. So, an equation like (\ref{eq:Taha}), cannot be self-consistently obtained in Rastall's model.

Moreover, since $T_{\mu \nu}$ is a perfect fluid and remembering the definition
of $S_{\mu \nu}$, Eq.(\ref{eq:conserv}), we get

\beq
\dot{\rho} + 3 H (\rho + p) = \frac{\kappa_{r} \lambda}{4\kappa_{r} \lambda - 1}
\left ( \dot{\rho} - 3 \dot{p} \right ) \ ,\
\label{eq: contrast}
\eeq
which generalizes the continuity equation for the Rastall's theory. 

It is worth noticing that, even without integrating the equations, one can immediately predict 
whether the universe expansion is accelerating or not by simply studying the sign of the right 
hand side of Eq.(\ref{eq:FR R1}).

Assuming for simplicity that the equation of state of the 
perfect fluid is a constant, i.e. setting $p = w \rho$, the condition $\ddot{a} > 0$ selects 
two possible regimes, the first one being 

\beq
w > \frac{6 \kappa_{r} \lambda - 1}{3(1 - 2 \kappa_{r} \lambda)} \ ,
\eeq
provided $\kappa_{r} \lambda > 1/4$. When $\lambda = 0$, however, the above relation
reduces to $w > -1/3$ in contrast with the GR result. We have therefore to choose the 
other solution, namely

\beq
w < \frac{6 \kappa_{r} \lambda - 1}{3(1 - 2 \kappa_{r} \lambda)} \ ,
\label{eq:EoS}
\eeq
with $\kappa_{r}\lambda < 1/4$. The right hand side of (\ref{eq:EoS}) may be positive or 
negative depending on the value of $\kappa_{r} \lambda$. More precisely, if $1/6 < \kappa_{r} 
\lambda < 1/4$, then the right hand side is positive, while it is negative for $0  < 
\kappa_{r} \lambda < 1/6$. It is worth stressing that, however, the model always gives a monotonic behaviour: always decelerated or, as in the above analyzed case, always accelerated. 

By the way, in the spirit of Cardassians, the only sources of gravity are radiation and matter. In particular, the recent epoch is driven by the matter content, described as a perfect fluid with equation of state $w=0$. With this constraint, it is easy to show that an accelerated behaviour is obtained for $\frac 1 6 < \kappa_{r}\lambda < \frac 1 4$, whereas we have a decelerated expansion choosing the following values $\kappa_{r}\lambda < \frac 1 6$ or $\kappa_{r}\lambda > \frac 1 4$.

\section{Rastall's model confronted with the data} \label{sec:data}

Neglecting the radiation component, the only fluid sourcing the gravitational
field is the standard matter, which can be modeled as dust, i.e. $p = 0$. In 
such a case, the continuity equation (\ref{eq: contrast}) is straightforwardly
integrated giving\,:

\begin{equation}
\rho \propto \rho_0 (1 + z)^{3 w_{eff}}\ ,\
\label{eq: rhorast}
\end{equation}
with a $0$ subscript denoting present day quantities, $z -1= \frac 1 a$ the redshift (having set $a_{0}=1$ for our flat-space universe), and 

\begin{equation}
w_{eff} = 1 - \frac{\kappa_r \lambda}{4 \kappa_r \lambda - 1}\ ,\
\end{equation}
an effective equation of state (EoS) for the dust matter, from which a relation between $w_{eff}$ and $\kappa_r \lambda$ is easily deduced. Note that, for $\lambda = 0$, 
one recovers the usual matter scaling $\rho \propto (1 + z)^3$, while
deviation from the standard behaviour occurs in $\lambda \neq 0$ Rastall's
theory. Such a different scaling is not surprising at all being an expected
consequence of the non-conservativity of the stress-energy tensor. Inserting
back Eq.(\ref{eq: rhorast}) into Eq.(\ref{eq: rf1}), we get

\begin{equation}
E^2 \doteq H^2/H_0^2 = (1 + z)^{3 w_{eff}}\ ,\
\label{eq: hubesq}
\end{equation}
which is all what we need to compute the luminosity distance

\begin{equation}
D_L(z, w_{eff}, h) = d_H (1 + z) \int_{0}^{z}{\frac{1}{E(z')} dz'}\ ,\
\label{eq: dl}
\end{equation}
with the Hubble radius $d_H = c/H_0 \simeq 3 h^{-1} {\rm \ Gpc}$
and $h$ the Hubble constant in units of $100 \ {\rm km/s Mpc}$. We have
now all the main ingredients to test the viability of the Rastall's model
by fitting the predicted luminosity distance to the data on the combined
Hubble diagram of SNeIa and GRBs. To this aim, we maximize the following
likelihood function:

\begin{equation}
{\cal{L}}(w_{eff}, h) \propto \exp{\left ( 
-\frac{\chi^2_{SNeIa} + \chi^2_{GRB}}{2} \right )}
\ \times \exp{\left [ - \left ( \frac{h_{HST} - h}{\sigma_{HST}}
\right )^2 \right ]}\ ,\
\label{eq: deflike}
\end{equation}
with

\begin{figure}[pt]
\begin{center}
\includegraphics [width=8.85cm,height=4.85cm]{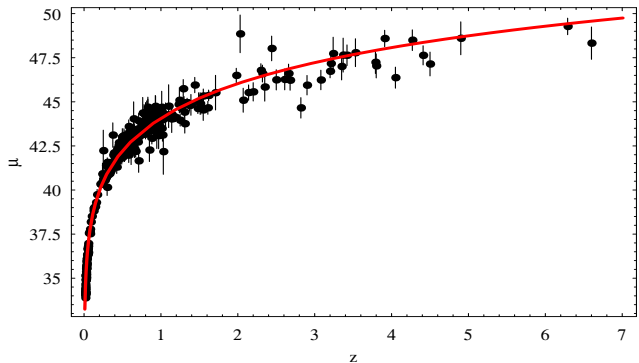}
\caption{Comparison among predicted and observed SNeIa and GRBs Hubble diagram.}
\label{fig: bestfit}
\end{center}
\end{figure}

\begin{equation}
\chi^2_{SNeIa} = \sum_{i = 1}^{{\cal{N}}_{SNeIa}}{\left [ 
\frac{\mu_{obs}(z_i) - \mu_{th}(z_i, w_{eff}, h)}{\sigma_i} \right ]^2} \ ,\
\label{eq: defchisneia}
\end{equation}

\begin{equation}
\chi^2_{GRB} = \sum_{i = 1}^{{\cal{N}}_{GRB}}{\left [ 
\frac{\mu_{obs}(z_i) - \mu_{th}(z_i, w_{eff}, h)}{\sigma_i} \right ]^2} \ .\
\label{eq: defchigrb}
\end{equation}
The $\chi^2$ terms in (\ref{eq: deflike}) take care of the Hubble 
diagram of SNeIa and GRBs, respectively, and rely on the distance 
modulus defined as

\begin{equation}
\mu_{th}(z, w_{eff}, h) = 25 + 5 \log{D_L(z, w_{eff}, h)} \ .\
\label{eq: defmuth}
\end{equation}
We use the Union \cite{union} dataset for SNeIa and the GRBs sample
assembled in Cardone et al. \cite{ccd09} to set the observed 
quantities $(\mu_{obs}, \sigma_i)$ for the
SNeIa and GRBs, respectively. Since the Hubble 
constant $h$ is degenerate with the (unconstrained) absolute magnitude
of a SN, we have added a Gaussian prior on $h$ using the results from the 
HST Key Project \cite{freedman} thus setting $(h_{HST}, \sigma_{HST}) 
= (0.72, 0.08)$. 

The best fit model turns out to be 

\begin{displaymath}
(w_{eff}, h) = (0.55, 0.68)\ ,\
\end{displaymath}
giving

\begin{displaymath}
\chi^2_{SNeIa}/d.o.f. = 1.08 \ , \ \ 
\chi^2_{GRB}/d.o.f. = 2.07 \ ,\
\end{displaymath}
where $d.o.f.= N_{SNeIa} + N_{GRB} - N_{p}$ is the number of degree of freedom of the model,
with $N_{SNeIa} = 307$ the number of SNeIa in the Union sample, $N_{GRB} = 69$ the number of GRBs, and $N_{p} = 2$ the number of parameters of the Rastall's model. While for SNeIa we get a very good reduced $\chi^2$, this is not the case for GRBs
so that one could be tempted to deem as unsuccessfull the fit. Actually, 
Fig.\,\ref{fig: bestfit} shows that the model is indeed fitting
quite well both the SNeIa and GRB data so that the large value of 
$\chi^2_{GRB}/d.o.f.$ should be imputed to the large scatter of the high
redshift data around the best fit line, not taken into account by the 
statistical error on the GRBs distance modulus. In order to further
test the model, one can consider the constraints on the matter density
parameter. Since our model only contains matter, one could naively think
that $\Omega_M = 1$. Actually, one must also take into account that 
$\Omega_M$ is defined using the GR coupling constant $\kappa_{GR}$ which 
is related to the Rastall's coupuling $\kappa_r$ through Eq.(\ref{eq: kappagrrast}).
It is then a matter of algebra to show that $\Omega_M = w_{eff}$ so that,
after marginalizing over $h$, we get the following constraints: 

\begin{displaymath} 
\Omega_M = 0.55_{-0.03 \ -0.05}^{+0.02 \ +0.05}\ ,
\end{displaymath} 
where we have used the notation $x^{+x_{1} +x_{2}}_{-y_{1} -y_{2}}$ to mean that $x$ is the median value of the parameter and $(x + x_{1}, x - y_{1}), (x + x_{2}, x - y_{2})$ are the 68\% and 95\% confidence ranges respectively. Note that the value thus obtained is in strong disagreement with the typical $\Omega_M \simeq 0.3$
obtained from both the cosmic microwave background radiation data and cluster gas mass fractions. Such a large
matter density parameter is clearly unacceptable and represents a strong
evidence against the Rastall's model. It is worth noting that such a 
result could be qualitatively foreseen considering that, because of 
the non-conservativity of the matter stress-energy tensor, a sort
of matter creation takes place thus increasing $\Omega_M$ and leading to 
the final disagreement.\\
Inverting the relation between $\Omega_M$ and $\kappa_r \lambda$, we get $\kappa_r \lambda = (1 - \Omega_M)/(3- 4 \Omega_M)$ so that, for $\Omega_M \simeq 0.3$, we get $\kappa_r \lambda \simeq 0.39 > 1/4$. Indeed, our best fit value for $w_{eff}$ gives back a value for $\kappa_r \lambda$ that falls outside the suitable range to reproduce an accelerated behaviour.

\section{Conclusions} \label{sec:conc}

In this paper we have focused on Rastall's theory of gravity,  which has been initially motivated by the need for a theory able to 
allow a non conservativity of the source stress-energy tensor without 
violating the Bianchi identities. In particular, we have reexamined this model of gravity to investigate the possibility that it could reproduce the observed cosmic speed up. First, we have explicitly worked out the modified Friedmann equations and  we have shown that 
Cardassian-like modifications of Friedmann equations are obtained in Rastall's model but, contrary to recent claims,  they cannot contain time-dependent parameters.
Then, we have confronted the model predictions with the available data concerning type Ia Supernovae (SNeIa) and Gamma Ray Bursts (GRBs):
what we have showed is that it is  possible to 
get an accelerated expansion that is in agreement with the Hubble diagram of 
both SNeIa and GRBs, even if there is unfortunately no possibility to reproduce an accelerated-decelerated-accelerated expansion for our universe as it seems to be requested. These results have also a major drawback: indeed, to get them it is necessary to postulate a matter density parameter much larger than the typical  $\Omega_M \simeq 0.3$ value inferred from cluster gas mass fraction data. As a consequence,  Rastall's theory is not in agreement with current cosmological observations of late time acceleration. However, since the non-conservativity of the matter stress-energy tensor can be related to matter creation, such a model could have important effects during the inflationary period, as it has been suggested \cite{smalley93}, even though an analysis of this  issue is beyond the scopes of the present paper.\\

\textbf{Note added. }After the publication of a preprint of this paper, the problem of structure formation in Rastall's theory has been studied in \cite{batista}, where the authors point out the difficulties of finding an agreement between this modified gravity model and the observational data.

\section*{Acknowledgments}

The authors warmly thank the attendants of the Journal Club on Extended and Alternative Theories of Gravity for useful discussions.
MC and VFC are supported by University of Torino 
and Regione Piemonte. Partial support from INFN projects PD51 and NA12 
is acknowledged too.

\end{document}